\documentclass[12pt]{article}

\title{
Critical Behavior of the Ferromagnetic Ising Model on a Sierpi\'nski Carpet:
Monte Carlo Renormalization Group Study
}
\author{Pai-Yi Hsiao{\small $^{(1)}$} and Pascal Monceau{\small $^{(1,2)}$}}
\begin{document}
\maketitle

\begin{center}
{\small $^{1}$Laboratoire de Physique Th\'{e}orique de la Mati\`{e}re
Condens\'{e}e,\\
P\^ole Mati\`ere et Syst\`eme Complexes (FR2438 CNRS), 
Universit\'{e} Paris 7 -- Denis~Diderot,\\
Case 7020, 2 Place Jussieu, 75251 Paris Cedex 05, France\\
$^{2}$D\'{e}partement de Physique et Mod\'{e}lisation, 
Universit\'{e} d'Evry-Val d'Essonne,\\
Boulevard Fran\c{c}ois Mitterrand, 91025 Evry Cedex, France }
\end{center}

\begin{abstract}

We perform a Monte Carlo Renormalization Group analysis of 
the critical behavior of the ferromagnetic Ising model on 
a Sierpi\'nski fractal with Hausdorff dimension $d_f\simeq 1.8928$.
This method is shown to be relevant to the calculation of the 
critical temperature $T_c$ and the magnetic eigen-exponent $y_h$ 
on such structures. On the other hand, scaling corrections hinder the 
calculation of the temperature eigen-exponent $y_t$. At last, the results 
are shown to be consistent with a finite size scaling analysis.

\end{abstract}
\noindent \textbf{PACS:}\newline
64.60.Ak Renormalization-group, fractal, and percolation studies of phase
transitions\\
75.10.Hk   Classical spin models\\
75.40.Mg   Numerical simulation studies\\
89.75.Da   Systems obeying scaling laws\\

\section{Introduction}
Since Mandelbrot attracted people's attention to the fractals 
in the 60's and 70's \cite{Mandelbrot75}, 
scientists began to take auto-similarity and scaling invariance 
as fundamental rules of nature.
In the description of second order phase transitions, 
Widom's homogeneity hypothesis \cite{Widom65} and 
Kadanoff's real space renormalization group scheme \cite{Kadanoff66} are 
based on the invariance of the physical behavior 
under any change of scale at the criticality  
where the correlation length is divergent. 
The validity of these hypothesis and scheme has been verified 
in different systems with an integer dimension which possess a translational symmetry 
\cite{Binder01}. As a matter of fact, a system with translational symmetry is 
auto-similar and can be considered as a particular case of fractal. 
Fractals are natural candidates to represent systems with non-integer 
dimensions. It turns out to be of fundamental relevance to know if 
the hypothesis and the scheme mentionned above still work on a 
{\it general} fractal system, where the translational symmetry is 
lost and replaced by scale invariance. 

The critical behavior of the Ising model on fractal lattices has been firstly
studied by Gefen and his co-workers in the 80's \cite{Gefen80s}. 
Their works were based on the Migdal-Kadanoff bond-moving renormalization method and 
showed that the topological features of the fractals plays an important role
in the determination of the critical behavior. As a main result, they found that 
a phase transition at non-zero temperature can occur only if the fractal has an 
infinite ramification order. Later, Bonnier {\it et al.} used an 
alternative decimation method \cite{Bonnier88} and high-temperature expansions
 \cite{Bonnier89} to study the phase transition on various 
fractals, namely Sierpi\'nski carpets. They also found that the critical exponents 
depend on the geometrical properties of the fractal.
In the above studies, the applied methods are approximative and 
the results are not always consistent. Moreover, in these theoretical analyses, 
the spins were placed at the corners of the occupied squares of the Sierpi\'nski carpets;
in these cases, the number of spins doesn't follow a power law of 
the lattice size and the Hausdorff dimension should not be expected to
enter in the description of the critical phenomena on such systems. 
Recently, due to the progress in the simulation methods and
the growth of the computer power, the critical behavior of the Ising model 
on fractals of Hausdorff dimension $d_f$ between 1 and 3 has been studied numerically 
in a much more precise way \cite{Monceau98, Carmona98, Hsiao00, Monceau01}.
The results showed that the finite-size scaling (FSS) analysis 
works in the case of fractals, although the convergence towards the thermodynamical limit
can be very slow when $d_f<2$, and that the hyperscaling law 
$d_f=2\beta/\nu+\gamma/\nu$ is verified. Moreover, discrepancies with the predictions 
of the $\epsilon$-expansions \cite{LeGuillou87} were observed. The universality of 
phase transitions on fractals is said to be weak \cite{Gefen80s, Wu87}. 
It is worth noting that the spins were placed 
at the center of the occupied squares of the Sierpi\'nski carpets in these studies; 
consequently, the number of the spins increases as a power law of the lattice 
size with an exponent equal to the Hausdorff dimension $d_f$.  
In this case, $d_f$ takes the place of the space dimension in 
Widom's homogeneity hypothesis because of the conservation of 
the total free energy under a change of length scale;
the generalization of this hypothesis in the case of fractals 
is quite straightforward \cite{Hsiao02}. 

Monte Carlo renormalization group (MCRG) has been shown to be 
a powerful tool in the study of critical phenomena 
\cite{Swendsen80s, Pawley84, Blote90s, Baillie92, Gupta96}. 
It combines Monte Carlo (MC) simulation techniques with the analysis 
of the real space renormalization group (RSRG). 
It has been, so far, applied to systems with translational symmetry.
A RSRG analysis involves an infinite number of couplings when dealing with the 
Hamiltonian. Theoreticians usually truncate the number of couplings to a finite one
in order to make the calculations tractable. 
It, therefore, makes RSRG an approximate method and the accuracy of 
the results is linked to the number of the couplings considered. 
Even if only few couplings are considered, the calculations associated with the 
process of decimation remain tedious.
With the aid of MC simulation, a renormalization-group calculation with 
a larger number of couplings becomes possible and the results 
are expected to be more reliable.However, some crucial problems are encountered in 
a MCRG study: 
\begin{enumerate}
\item[(1)] The critical slowing down:\\
Hypothetically, the accuracy of the thermodynamical averages can be improved 
by performing a larger number of MC steps to reduce the statistical errors.
In fact, due to the divergence of the correlation length at the criticality,
local MC algorithms (for e.g., the Metropolis one) 
suffer from critical slowing down, which hinders the computation of
accurate thermodynamical averages. Fortunately, the use of 
cluster algorithms, for e.g., the Swendsen-Wang algorithm \cite{Swendsen87} 
and the Wolff algorithm \cite{Wolff89},
enables to improve significantly the efficiency of the simulations.

\item[(2)] The finite size effects:\\
RSRG method involves an infinite size lattice,
whereas computer simulations can only be performed 
on finite size lattices. One should perform the simulation 
on several large systems and then extrapolate the results to 
the infinite one. \end{enumerate}
The MCRG method, according to which one plays the game of changing the length scale,
should be a natural approach to explore the critical behavior 
of phase transitions on hierarchical lattices.     
The purpose of this paper is to study the critical behavior of the Ising model
on a Sierpi\'nski carpet using the MCRG method. As far as we know, it is the first 
time that the MCRG method is used in the case of fractal lattices. 
The paper is organized as follows : In Sec.~II, we present the model and 
recall briefly the MCRG method. We explain how the simulation is setup in Sec.~III.
Sec.~IV is devoted to the calculation of the critical temperature $T_c$, 
the temperature eigen-exponent $y_t$, and the magnetic eigen-exponent $y_h$. 
A FSS analysis is set out in Sec.~V for a 
consistency check. 

\section{Sierpi\'nski carpet and MCRG method}
A 3 by 3 square lattice with its central subsquare removed
is chosen as ``generating cell''.
The Sierpi\'nski carpet we deal with is constructed iteratively 
from this generating cell:
\begin{enumerate}
\item
We take the generating cell as the carpet at the first iteration step
and denote it $SC(3,1,1)$.
\item
The carpet at the $(k+1)$-th iteration step $SC(3,1,k+1)$ 
is constructed by enlarging three times the size of 
the carpet at the $k$-th iteration step and replacing each enlarged 
un-removed subsquare by the whole generating cell. This carpet 
becomes a ``true'' fractal in the mathematical sense when $k$ tends to infinity.
\end{enumerate}
The size of the lattice $SC(3,1,k)$ is equal to $L_k=3^k$. 
A spin is set at each center of an un-removed subsquare (so-called occupied square).
The number of spins on $SC(3,1,k)$ is therefore equal to $N_k=8^k$. We notice that 
$N_k$ is equal to ${L_k}^{d_f}$ where $d_f=\log 8/\log 3 \simeq 1.892789$ is the Hausdorff dimension.
The critical behavior of the ferromagnetic Ising model on this Sierpi\'nski carpet
has been recently studied independently by P.~Monceau {\it et al.} 
\cite{Monceau98, Monceau01} and J.~Carmona {\it et al.} \cite{Carmona98} 
by means of MC simulation. Thus, estimations of the critical temperature are available 
and the comparison between the two methods will be possible. 

Let us briefly recall the MCRG method.The geometrical structure of a 
Sierpi\'nski carpet remains invariant during 
a RSRG process involving a change in the length scale from 1 to $b$, 
provided that $b$ is equal to an integer power of the size of its generating cell.
In our case, we set $b$ equal to 3 and increase the length scale by this factor 
at each renormalization step.
The majority rule is used to decimate the spin blocks:
a new spin is assigned to a given block according to the sign of the
summation of the spin states of the 8 occupied sites in this block;
if the sum is zero, a spin state $+1$ or $-1$ is assigned to the block 
with the same probability. The reduced Hamiltonian after $n$ renormalization steps 
($n$ is a positive integer) reads as
\begin{equation}
{\cal H}^{(n)}(\{K_\alpha^{(n)}\}_{\alpha=1}^{\infty})
=\sum_{\alpha=1}^{\infty}\,K_\alpha^{(n)}S_\alpha^{(n)}\ ,
\end{equation} 
where $\{K_\alpha^{(n)}\}_{\alpha=1}^{\infty}$ is the coupling set 
after $n$ renormalization steps 
and $S_{\alpha}^{(n)}$'s are the conjugate lattice sums of the spin products 
on the reduced network.
The matrix $\bf{\cal T}$, which describes the flow of the couplings
from the $n$-th renormalization step to the $(n+1)$-th one,
is defined by 
\begin{equation}
{\cal T}^{(n+1, n)}_{\alpha\beta} = 
\frac{\partial K^{(n+1)}_{\alpha}}{\partial K^{(n)}_{\beta}} 
\end{equation}
and satisfies the following relation: 
\begin{equation}
\left[\partial S \right]^{(n+1,n)}_{\gamma\beta} = \sum_{\alpha=1}^{\infty}\,
\left[\partial S \right]^{(n+1,n+1)}_{\gamma\alpha}\, 
{\cal T}^{(n+1, n)}_{\alpha\beta}, 
\label{S_ST}
\end{equation} 
where 
$\left[\partial S \right]^{(n,m)}_{\alpha\beta} \equiv   
\partial \langle S^{(n)}_{\alpha}\rangle/\partial K^{(m)}_{\beta}$
can be calculated by 
$\langle S^{(n)}_{\alpha} S^{(m)}_{\beta}\rangle - 
\langle S^{(n)}_{\alpha}\rangle \langle S^{(m)}_{\beta}\rangle$.
One can hence calculate the matrix ${\cal T}$ by inverting the above relation.
Two critical exponents enable to describe the static scaling 
properties of the system:  the temperature eigen-exponent $y_t$ 
and the magnetic one $y_h$.
They describe respectively the scaling behavior of 
the reduced temperature $t=T/T_c-1$ (where $T_c$ is the critical temperature) 
and the external magnetic field $h$ near the criticality
under a change of the length unit from $1$ to $b$:
\begin{eqnarray}
\nonumber t &\to& t'= b^{y_t}t\ ,\\
\nonumber  h &\to& h'= b^{y_h}h\ .
\end{eqnarray} 
$b^{y_t}$ and $b^{y_h}$ are associated with the two relevant directions of
the renormalization flows and can be obtained by finding the largest eigenvalues 
of the ${\cal T}$ matrix in the even- and odd-coupling subspaces, respectively.

\section{Simulation setup}
Up to 4-spin couplings have been considered in our study and the interaction 
range has been restricted within a $3 \times 3$ square.
There are 25 even couplings and 11 odd couplings
as shown in Fig.1(a) and Fig.1(b), respectively.
The symmetry number obtained by rotating and reflecting 
a given coupling is indicated in the third column of the figure. 
These couplings are listed in decreasing order 
according to the importance factor proposed by Bl\"ote {\it et al.} 
\cite{Blote90s}: 
$F=(2^{s/2} \bar{r})^{-1}$, 
where $\bar{r}$ is the average distance between the $s$ spins.
The value of $F$ is given in the right column of the figure. 
The simulations have been carried out at two different temperatures,  
$T=1.4813$ obtained by Carmona {\it et al.} \cite{Carmona98} 
and $T=1.4795$ obtained by  Monceau {\it et al.} \cite{Monceau01}. 
The size of the starting Sierpi\'nski carpet 
varies between $3^4=81$ and $3^8=6561$. 
The simulation procedure is organized as follows:
\begin{enumerate}
\item 
An equilibrium spin configuration on the starting Sierpi\'nski carpet $SC(3,1,k)$ 
is generated by the Wolff algorithm \cite{Wolff89} at the simulation temperature. 
\item 
The lattice sums conjugated to the 25 even couplings
and to the 11 odd ones are calculated.
\item
A decimation is done, according to the majority rule, 
by dividing the size of the Sierpi\'nski carpet by 3 and 
the lattice sums associated with the renormalized carpet are calculated.
\item
The step 3 is repeated until the size of the renormalized Sierpi\'nski carpet 
is smaller than 9.
\end{enumerate}

A data sample of the lattice sums is built up from 
$10^5$ steps of the above simulation, with periodic boundary conditions.
At each simulation temperature, for the five 
different sizes of the starting Sierpi\'nski carpet, 
10 independent data samples are collected. 
The errors bars are estimated from a standard statistical analysis of these data.

\section{MCRG Results}
\subsection{Critical temperature $T_c$}
The size of the lattice after $n+p$ renormalization steps  
from the starting Sierpi\'nski carpet of size $L_{k+p}$ is equal to 
the size of the lattice after $n$ renormalization steps 
from the starting Sierpi\'nski carpet of size $L_{k}$. 
At the fixed point $\bf{K^{*}}$, the thermodynamical averages of 
the lattice sums calculated from renormalized carpets with the same 
size should be independent of the size of the starting Sierpi\'nski carpet. 
Let us assume that the initial coupling vector $\bf{K^{(0)}}$ 
is close to $\bf{K^{*}}$. The difference of the lattice sums on 
the reduced lattices of equal size, 
derived from the small displacement $\delta \bf{K}=\bf{K^{*}}-\bf{K^{(0)}}$,  
should satisfy the following relation:  
\begin{equation}
\left. \left(
 \langle S^{(n+p)}_\alpha \rangle_{L_{k+p}}
-\langle S^{(n)}_\alpha \rangle_{L_{k}}
\right) \right|_{\bf{K^{(0)}}}
\simeq 
 -\sum_{\beta}\, 
 \left. \left(\left[\partial S_{L_{k+p}}\right]^{(n+p,0)}_{\alpha\beta} 
 -\left[\partial S_{L_{k}} \right]^{(n,0)}_{\alpha\beta}\right)
  \right|_{\bf{K^{(0)}}}\, \delta K_{\beta} ,
\label{fixedpoint}
\end{equation}
where the suffix $L_{k+p}$ or $L_{k}$ indicates that the physical quantity is
obtained from the starting Sierpi\'nski carpet of size $L_{k+p}$ or $L_{k}$.
In the even coupling subspace, the nearest neighborhood coupling $K_1$
represents the inverse of the temperature.
If we neglect the contribution from the directions other than
the nearest neighborhood coupling and 
set $\alpha=1$ in Eq.(\ref{fixedpoint}), 
the critical temperature can be calculated by 
\begin{eqnarray}
\nonumber
T_c^{[k+p,k]}(n) &=& \left( K^*_1 \right)^{-1} 
=\left[ K_1^{(0)} +\delta K_1\right]^{-1}\\ 
&\simeq& \left[ K_1^{(0)} -
\frac{\left. \left(
  \langle S^{(n+p)}_1 \rangle_{L_{k+p}}-\langle S^{(n)}_1\rangle_{L_{k}}
  \right) \right|_{K_1^{(0)}}}
{\left. \left(\left[\partial S_{L_{k+p}} \right]^{(n+p,0)}_{1\ 1}
 -\left[\partial S_{L_{k}} \right]^{(n,0)}_{1\ 1}\right)
  \right|_{K_1^{(0)}}}
 \right]^{-1}\ , 
\label{Tc}
\end{eqnarray}
where the symbol $T_c^{[k+p,k]}(n)$ indicates that the temperature is obtained
from two starting carpets of size $3^{k+p}$ and $3^{k}$ after
$n+p$  and $n$ renormalization steps respectively.
Fig.~2 shows the evolution of
$T_c^{[k+p,k]}(n)$ calculated from pairs of the starting Sierpi\'nski carpets 
at the simulation temperatures 
$T_{sim}= \left[K_1^{(0)}\right]^{-1} = 1.4813$ and 
$T_{sim}= \left[K_1^{(0)}\right]^{-1} = 1.4795$.  
These results deserve the following comments :

i) At the two simulation temperatures,  $T_c^{[k+p,k]}(n)$ decreases as $n$ increases and
tends to converge to some value. 

ii) If we fix $p=1$, the value of $T_c^{[k+p,k]}(n)$ decreases as 
$k$ increases. We can see that $T_c^{[5,4]}(n) > T_c^{[6,5]}(n) >T_c^{[7,6]}(n) >T_c^{[8,7]}(n)$ and 
the curves $T_c^{[7,6]}(n)$ and $T_c^{[8,7]}(n)$ are very close to each other.

iii) $T_c^{[k+p,k]}(n)$ decreases as $k$ decreases
if the size of the largest starting Sierpi\'nski carpet $3^{k+p}$ is fixed. One can observe that 
$T_c^{[8,7]}(n)>T_c^{[8,6]}(n)>T_c^{[8,5]}(n)>T_c^{[8,4]}(n)$
where the size of the largest starting carpet is fixed at 6561.
These curves tend to converge to some temperature while $n$ increases.

The best estimation for the critical temperature can be obtained from 
the 6-data points $T_c^{[8,7]}(n)$ for $n=0,1,\cdots,5$,
by performing the following three-parameter $(T_c, A_0, n_0)$ fit:
\begin{equation}
T_c^{[k+p,k]}(n) = T_c + A_0 3^{-n /n_0} \ .
\end{equation} 
The results of the fits at the two simulation temperatures lead to : 
\begin{eqnarray}
\nonumber
\mbox{at $T_{sim}=1.4813$,} & &  
(T_c,\, A_0,\, n_0)=(1.47959(15),\, 0.02718(28),\, 0.839(25))\\ 
\nonumber
& & \mbox{with the reliability $R^2=0.99968$ and}\\ 
\nonumber
\mbox{at $T_{sim}=1.4795$,} & &  
(T_c,\, A_0,\, n_0)=(1.47933(27),\, 0.02827(49),\, 0.868(42))\\
\nonumber   
& & \mbox{with the reliability $R^2=0.99911$}\ . 
\end{eqnarray}
The best value of  the critical temperature can be estimated by taking the average of the 
results of the two fits; we hence have $T_c=1.47946(16)$.

In the following subsections, the eigen-exponents will be calculated 
from simulations carried out at $T_{sim}=1.4795$, since 
it lies closer to the best estimate of the critical temperature $T_c=1.47946(16)$. 
The simulations carried out at $T_{sim}=1.4813$ will provide a consistency check.
It is well known that the exponents provided by the MCRG method 
are not very sensitive to small deviations from the critical temperature.

\subsection{Temperature eigen-exponent $y_t$}
\label{subsec_y_t}
We work in the even-coupling subspace and 
consider in turn the first 1 to 25 couplings shown in Fig. 1(a). 
We calculate the $\bf{\cal T}$ matrix in the subspace 
at each renormalization level by inverting Eq.(\ref{S_ST}) with the help 
of the Gauss-Jordan elimination method
($\S 2.1$ in Ref.\cite{NR}).
The dimension of $\bf{\cal T}$ is $N_{coupling} \times N_{coupling}$
where $N_{coupling}$ varies between 1 and 25. 
The largest eigenvalue $\lambda_t$ of $\bf{\cal T}$ is obtained by applying 
the eigenvalue searching method for a real non-symmetric square matrix
given in $\S 11.5$ and $\S 11.6$ of Ref.\cite{NR}. 
The temperature eigen-exponent is then calculated by $y_t=\log_3 \lambda_t$.
Fig.~3 shows the evolution of the eigen-exponent $y_t$ 
as a function of the number of  even couplings $N_{coupling}$,
obtained from starting Sierpi\'nski carpets of different sizes
and at different renormalization levels.  
One can see that $y_t$ tends to converge towards stable values when 
$N_{coupling}$ increases, within the statistical errors, 
excepted at the highest renormalization levels where the sizes 
of the reduced lattices are $3^3$ and $3^2$.
We estimate $y_t$ in the infinite couplings limit by taking 
the average of the values of $y_t$ obtained from 16 to 25 even couplings. 
The average value at each renormalization level is reported 
in Table \ref{tab_yt} where we have disregarded the unreliable results 
at the highest renormalization levels. 
We find that the finite-size effect is not significant
because at the same renormalization level, the values of  
$y_t$ obtained from the starting Sierpi\'nski carpets of different sizes 
are about the same. 
A similar situation has been observed in the study of  the 3D Ising model on regular lattices 
\cite{Blote90s, Baillie92}.
In order to extrapolate the value of $y_t$ on the starting Sierpi\'nski
carpet of infinite size,
we plot in the Fig.~4 the value of $\lambda_t$ obtained 
at level $n-(n+1)$ from the starting carpet $SC(3,1,n+4)$ 
with respect to $n$.
We find that, up to the renormalization levels which were performed, 
$\lambda_t$ seems not to be convergent.
This behavior is quite different from the one observed in the case 
of  the 3D Ising  model on regular lattices 
\cite{Blote90s, Baillie92} where $\lambda_t$ 
tends to converge to some value while $n$ increases. 
We, hence, cannot estimate the convergence by performing the three-parameter fit
($\lambda_t^*, a_t, \omega_t$) proposed in Ref.\cite{Baillie92}:
\begin{equation}
\lambda_t = \lambda^*_t + a_t 3^{-\omega_t n}\ .
\end{equation}
The lack of convergence in the case of the Sierpi\'nski carpet 
may be due to a small value of the scaling correction exponent 
$\omega_t$ . At the present stage, we are only able to provide an upper bound 
for the Sierpi\'nski carpet of infinite size:
$\lambda_t<1.781(10)$ or $y_t <0.5254(51)$.
The same situation is also observed at $T_{sim}=1.4813$.
The difficulty in estimating the value of $y_t$ has already been observed
in the study based on FSS analysis \cite{Monceau01}.
According to Widom's homogeneity hypothesis, 
$y_t$ is equal to the inverse of the correlation-length exponent $\nu$.
Monceau et al. \cite{Monceau01} have shown that 
the maxima values of the logarithmic derivatives
$\Phi_i$ (for $i=1,2$), defined by 
\begin{equation}
\Phi_i = \langle E\rangle -\frac{\langle E|M|^i\rangle}{\langle |M|^i\rangle} 
\end{equation}
where $E$ is the total energy and $M$ the total magnetic moment,
are affected by scaling corrections.
Instead of scaling {\it properly} as a power law,
the maxima values of $\Phi_i$ (for $i=1,2$) appear to exhibit a slight concavity 
in a log-log plot with respect to the lattice size.
Monceau et al. could only provide an upper bound for $y_t$ from their MC study: 
$1/\nu <1/1.565 \simeq 0.639$. Our estimation is consistent with their result.

\subsection{Magnetic eigen-exponent $y_h$}
\label{subsec_y_h}
We now focus on the odd-coupling subspace.
We consider in turn the first $N_{coupling}$ odd couplings shown in 
Fig.1(b) where $N_{coupling}= 1, 2, \dots, 11$.
The ${\cal T}$ matrix in the odd-coupling subspace is obtained from
Eq.(\ref{S_ST}) by considering only the lattice sums conjugated to
the odd couplings.
We denote $\lambda_h$  the largest eigenvalue of this matrix, and 
the magnetic eigen-exponent is calculated by $y_h=\log_3 \lambda_h$. 
Fig.~5 shows the evolution of $y_h$ as a funtion of the number
of odd couplings at each renormalization level
obtained from the starting Sierpi\'nski carpet $SC(3,1,k)$ with $k=4,5,\cdots,8$.
We can see that $y_h$ tends to be stable as $N_{coupling}$ increases.
Larger fluctuations are observed at the highest renormalization level 
where $y_h$ is extracted from the reduced carpets of sizes $3^3$ and $3^2$;
we will disregard these values. 
We estimate $y_h$ in the infinite couplings limit by taking the average value
of them when 7 to 11 odd couplings are considered.
The results are reported in Table \ref{tab_yh}.
The size effect of the starting Sierpi\'nski carpet 
on the evolution of $y_h$ with respect to the renormalization level 
is more significant than it is on the evolution of $y_t$. 
However, the value of $y_h$ at level $n-(n+1)$ from the starting
carpet $SC(3,1,n+4)$ tends to converge to some value  
as $n$ increases (see Fig. 6). 
We hence perform the three-parameter fit 
suggested by Baillie {\it et al.} \cite{Baillie92}:
\begin{equation}
\lambda_h = \lambda_h^* + a_h 3^{-\omega_h n}\ . 
\end{equation} 
The fit yields  
$(\lambda_h^*,\,a_h,\,\omega_h)= (7.38298(66),\, -0.0737(12),\, 1.86(12))$
with a reliability equal to $R^2=0.99949$.
According to these results, the magnetic eigen-exponent $y_h$ takes the value $1.81973(9)$
and the associated scaling correction exponent $\omega_h$ the value $1.86(12)$. 
$y_h$ is slightly larger than the one obtained by Monceau et al. \cite{Monceau01}.
They found that $\gamma/\nu=1.732(4)$ and $\beta/\nu = 0.075(10)$. 
According to the relations: 
$\gamma/\nu=2y_h-d_f$ and $\beta/\nu=d_f-y_h$ \cite{Hsiao02},
their results correspond to $y_h=1.812(2)$ and $y_h=1.818(10)$, respectively.
No significant discrepancy can be brought out from these results, since the relative 
difference between the exponent provided respectively by the MCRG method and MC simulations 
remains smaller than $0.5$ percent.

\section{Results from FSS analysis}
\label{FSSR}
The lattice sums associated with the first odd and the first even couplings 
represent respectively the total magnetic moment $M$ and 
minus the total energy $E$. 
We can, hence, perform  a FSS analysis for a consistency check.  
According to this analysis, the thermodynamical average of 
the total magnetic moment at $T_c$ 
should scale as $\langle |M| \rangle \sim L^{d_f- \beta/\nu} = L^{y_h}$
and the derivative of the Binder's cumulant 
$U=1-\langle M^4 \rangle /(3\langle M^2 \rangle^2)$  at the critical temperature $T_c$ scales as
\begin{equation}
\frac{d\,U}{d\,\beta_B} = -(1-U)(\Phi_4- 2\Phi_2) \sim L^{1/\nu}= L^{y_t} \ ,
\label{dU}
\end{equation} 
where $L$ is the lattice size and
$\beta_B = (k_BT)^{-1}$ is the inverse of temperature. 
In Fig.~7, we plot the values of $\langle |M|\rangle$ and 
${d\,U}/{d\,\beta_B}$ at $T=1.4795$  
for the 5 different lattice sizes from $3^4$ to $3^8$ in logarithmic coordinates.
We find that the values of $\langle |M|\rangle$ line up along straight lines, 
with the lattice size covering several order of magnitude.
The slope provided by a least-square fit from the $5$ above points,  
is $1.8198(11)$, with a reliability $R^2=0.99994$.
It turns out that this value is consistent with the value of $y_h$ obtained 
from MCRG method in the subsection \ref{subsec_y_h}.
The behavior of  ${d\,U}/{d\,\beta_B}$ is not affected by scaling corrections, 
(see Fig.7), although the $\Phi_i$ (for $i=1,2$) 
exhibit a slight concavity as a function of the lattice size in a log-log plot.
The reason why ${d\,U}/{d\,\beta_B}$ is not affected by scaling corrections may be
a kind of ``magic'' cancellation of these scaling correction effects 
in the difference between $\Phi_4$ and $2\Phi_2$ .
The slope measured from ${d\,U}/{d\,\beta_B}$ is $0.449(6)$ 
with the fitting reliability $R^2=0.99948$.
This value is consistent with the result obtained in the subsection
\ref{subsec_y_t} where $y_t<0.5254(51)$. 

\section{Conclusion}
\label{CONCL}

The MCRG method has been shown to 
provide reliable values of the critical temperature $T_c$ and 
the magnetic eigen-exponent $y_h$, for the Ising model in a case where the 
underlying network has a fractal structure. The scaling correction exponent 
$\omega_h$ associated with $y_h$ can also be calculated. 
Moreover, difficulties are encountered in the estimation of 
the temperature eigen-exponent $y_t$ where, unlike regular lattices, 
the convergence of the renormalization flow in this direction 
seems to be very slow; nevertheless, an upper bound for $y_t$ can be provided. 
A similar slowness in the convergence of the logarithmic derivatives 
of the magnetization $\Phi_i$ to the thermodynamical limit has already 
been brought out from MC simulations in the case of the same 
Sierpi\'nski fractal \cite{Carmona98} \cite{Monceau01}. 
In reference $11$, scaling corrections have been shown
to affect strongly the behavior of the $\Phi_i$'s and only an upper bound 
for $y_t$ has been calculated from their finite size behavior. 
The difficulties in calculating $y_t$ for fractal structures 
arise in the two methods, but emerge differently. At last, the results provided 
by the MCRG method have been shown to be consistent with the 
one obtained by Monceau et al. \cite{Monceau01} and with an additional 
FSS analysis of the MCRG results.    

\section*{Acknowledgements}
A part of the numerical simulations has been carried out in the 
Institut de D\'eveloppement et des Ressources en Informatique Scientifique 
(IDRIS), supported by the Centre National de la Recherche Scientifique 
(project number $021186$). We acknowledge the scientific committee and 
the staff of the center. We are also grateful to the Centre de Calcul Recherche 
(CCR) of the University Paris VII--Denis Diderot, 
where the rest of the simulations have been done.


\clearpage 
\section*{Figure captions}
\begin{enumerate}
\item[Fig.1]  
(a) 25 even couplings and (b) 11 odd couplings, considered in our MCRG study.
They are ordered according to the Bl\"ote's importance factor $F$
whose value is given in the right colomn.
The symmetry number is given in the next to last right colomn.

\item[Fig.2]
Evolution of $T_c^{[k+p,k]}(n)$ with respect to the different pairs of
the starting Sierpi\'nski carpets $SC(3,1,k+p)$ and $SC(3,1,k)$
(denoted by $[k+p,k]$) obtained at 
(a) $T_{sim}=1.4813$, (b) $T_{sim}=1.4795$. 

\item[Fig.3]
Variation of $y_t$ with respect to the number of even couplings $N_{coupling}$
at different renormalization levels from the starting Sierpi\'nski carpet
$SC(3,1,k)$ with $k=4,5,\cdots,8$.

\item[Fig.4]
$\lambda_t$ obtained at level $n-(n+1)$ from the starting carpet 
$SC(3,1,n+4)$.

\item[Fig.5]
Variation of $y_h$ with respect to the number of odd couplings $N_{coupling}$
at different renormalization levels from the starting Sierpi\'nski carpet
$SC(3,1,k)$ with $k=4,5,\cdots,8$.

\item[Fig.6]
$\lambda_h$ obtained at level $n-(n+1)$ from the starting carpet 
$SC(3,1,n+4)$.

\item[Fig.7]
$\langle |M|\rangle$ and ${d\,U}/{d\,\beta_B}$ at $T=1.4795$
on SC(3,1,k) with $k=4,5,\cdots,8$ in logarithmic coordinates.                           
\end{enumerate}

\clearpage 
\begin{table}
\caption{
Average value of $y_t$ calculated from the starting Sierpi\'nski carpets
         at different renormalization levels.
         It is obtained by considering $16$ to $25$ even couplings
         at the simulation temperature $T_{sim}=1.4795$. 
         At the highest renormalization level, these average value are disregarded
         (denoted by the symbol ``---'') 
         because the eigen-exponent is not stable with respect to 
         the number of the even couplings. 
}
\label{tab_yt}
\begin{center}
\begin{tabular}{|c|ccccc|}
\hline
level &$SC(3,1,8)$&$SC(3,1,7)$&$SC(3,1,6)$&$SC(3,1,5)$&$SC(3,1,4)$\\ 
\hline
$0-1$ & 0.7497(17)& 0.7437(20)& 0.7474(14)& 0.7506(26)& 0.7483(26)\\
$1-2$ & 0.7186(27)& 0.7235(14)& 0.7225(19)& 0.7221(17)&   ---     \\
$2-3$ & 0.6465(26)& 0.6459(32)& 0.6466(25)&   ---     &           \\
$3-4$ & 0.5807(19)& 0.5883(23)&   ---     &           &           \\
$4-5$ & 0.5254(51)&   ---     &           &           &           \\
$5-6$ &   ---     &           &           &           &           \\
\hline
\end{tabular}
\end{center}
\end{table}

\begin{table}
\caption{
Average value of $y_h$ calculated from the starting Sierpi\'nski carpets
         at different renormalization levels.
         It is obtained by considering $7$ to $11$ odd couplings
         at the simulation temperature $T_{sim}=1.4795$. 
         At the highest renormalization level, these average values are disregarded
         (denoted by the symbol ``---'') 
         because the eigen-exponent is not stable with respect to 
         the number of the odd couplings. 
}
\label{tab_yh}
\begin{center}
\begin{tabular}{|c|ccccc|}
\hline
level &$SC(3,1,8)$&$SC(3,1,7)$&$SC(3,1,6)$&$SC(3,1,5)$&$SC(3,1,4)$\\ 
\hline
$0-1$ & 1.80753(8)& 1.80762(3)& 1.80799(4)& 1.80861(5)& 1.81060(8)\\
$1-2$ & 1.81602(2)& 1.81621(3)& 1.81673(3)& 1.81857(4)&   ---     \\
$2-3$ & 1.81713(1)& 1.81759(3)& 1.81948(4)&   ---     &           \\
$3-4$ & 1.81807(1)& 1.81985(2)&   ---     &           &           \\
$4-5$ & 1.81967(3)&   ---     &           &           &           \\
$5-6$ &   ---     &           &           &           &           \\
\hline
\end{tabular}
\end{center}
\end{table}

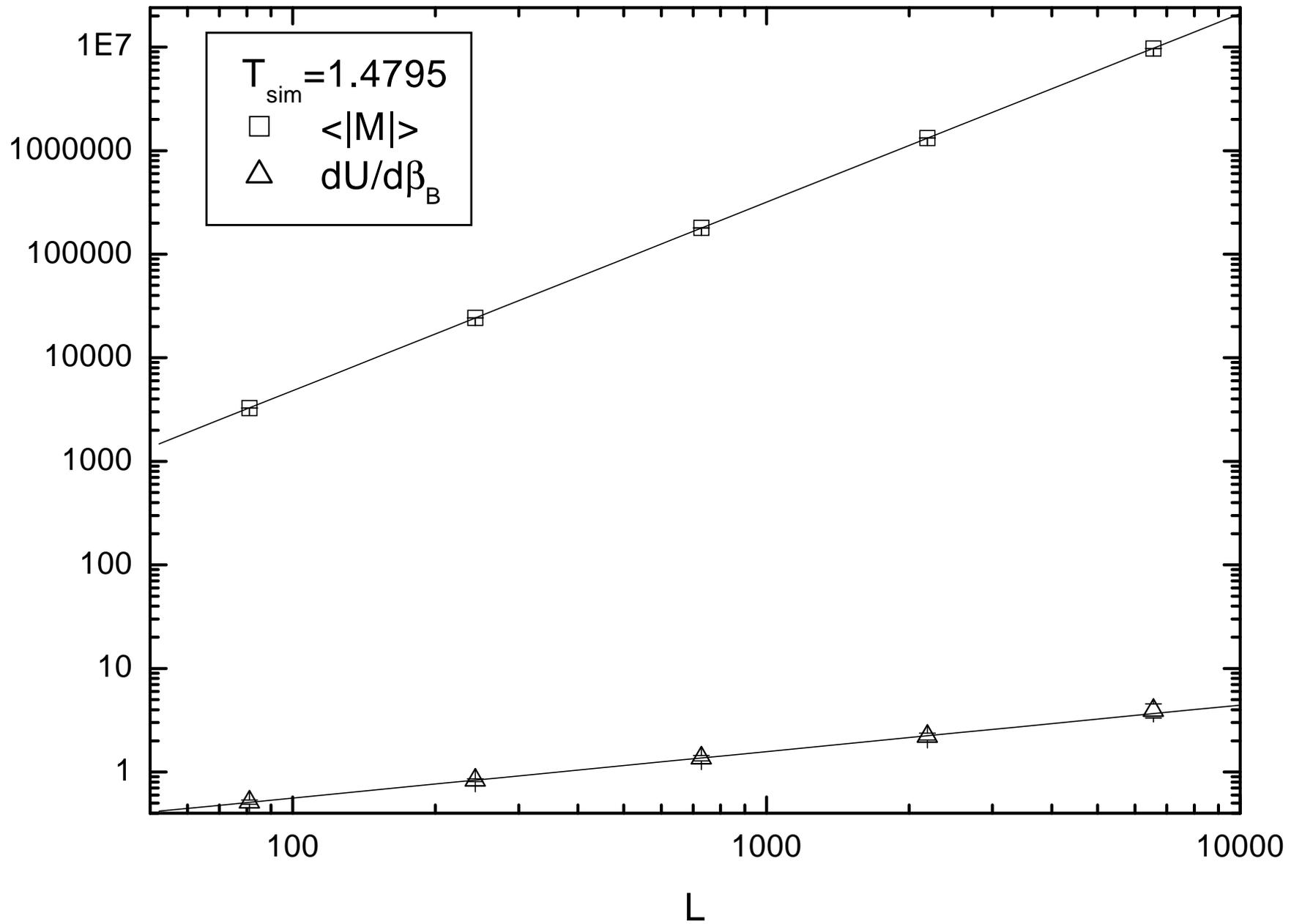
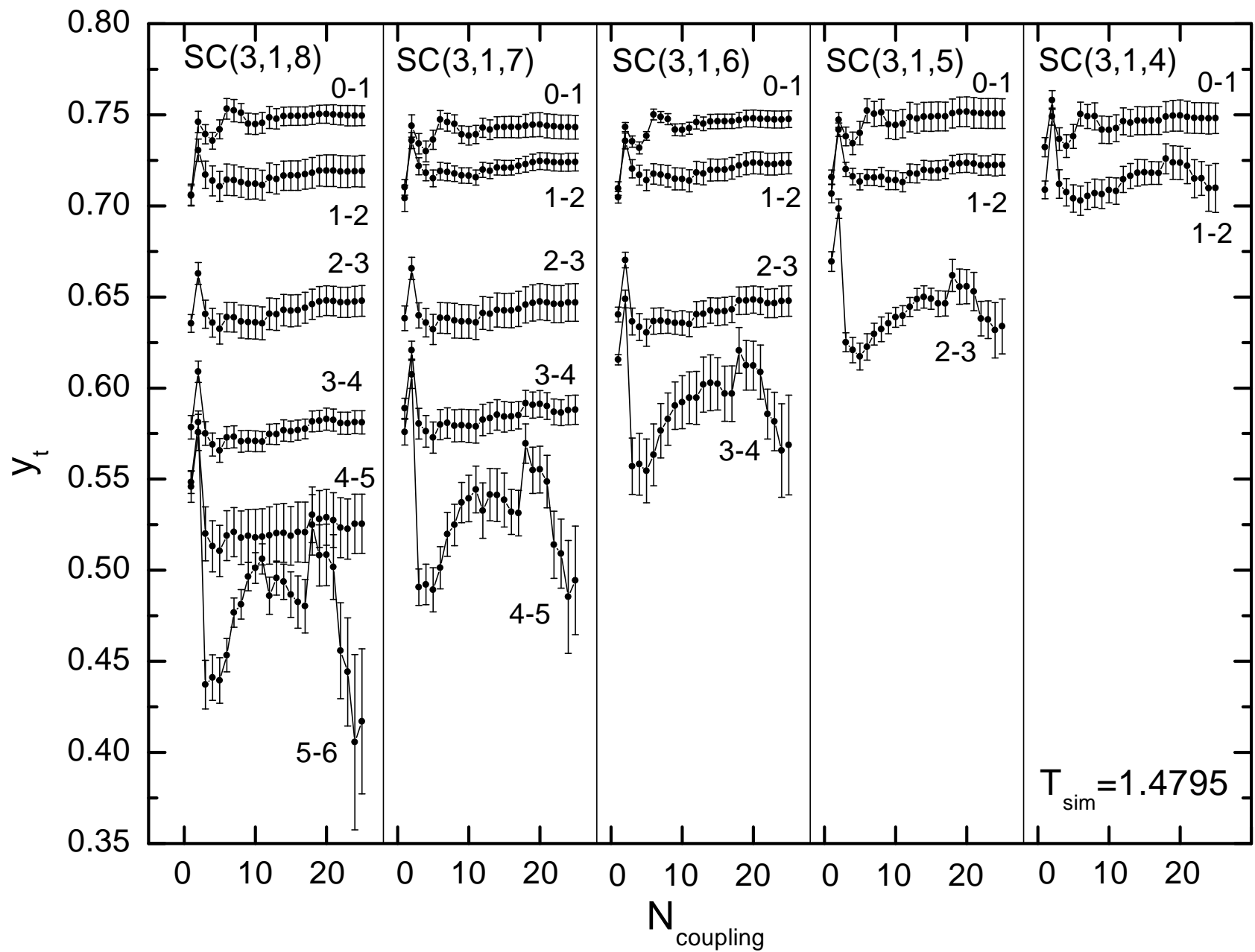
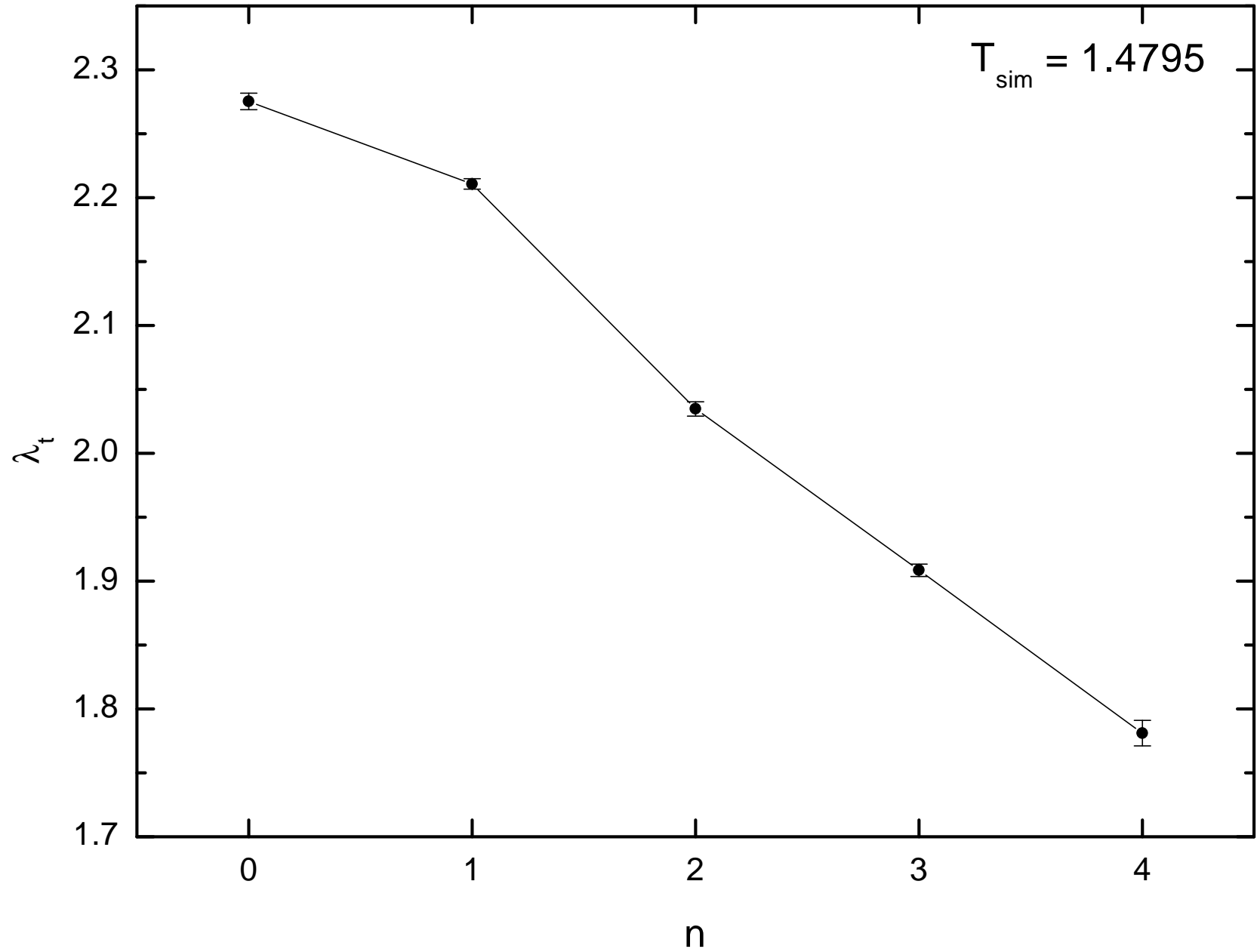
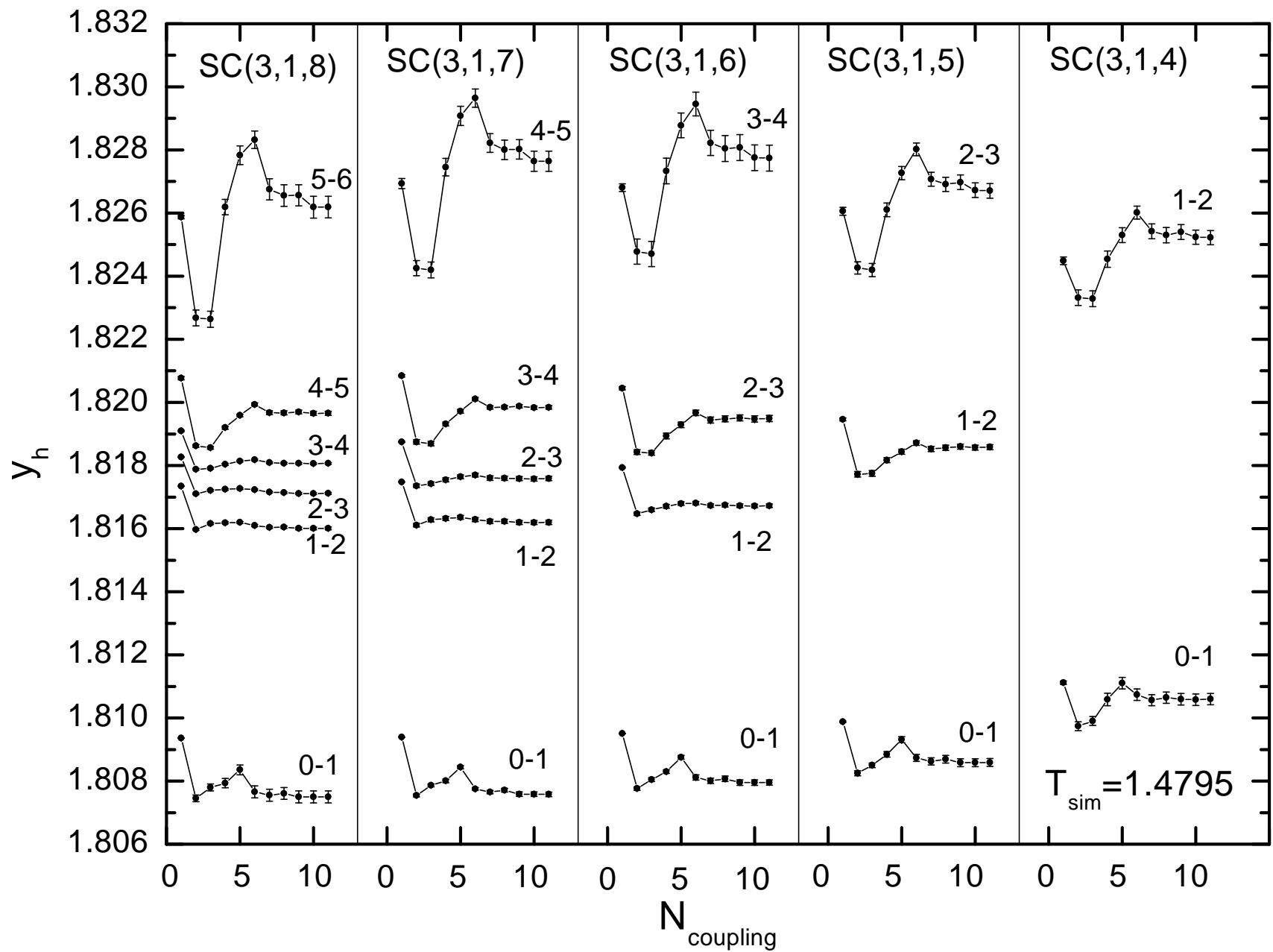
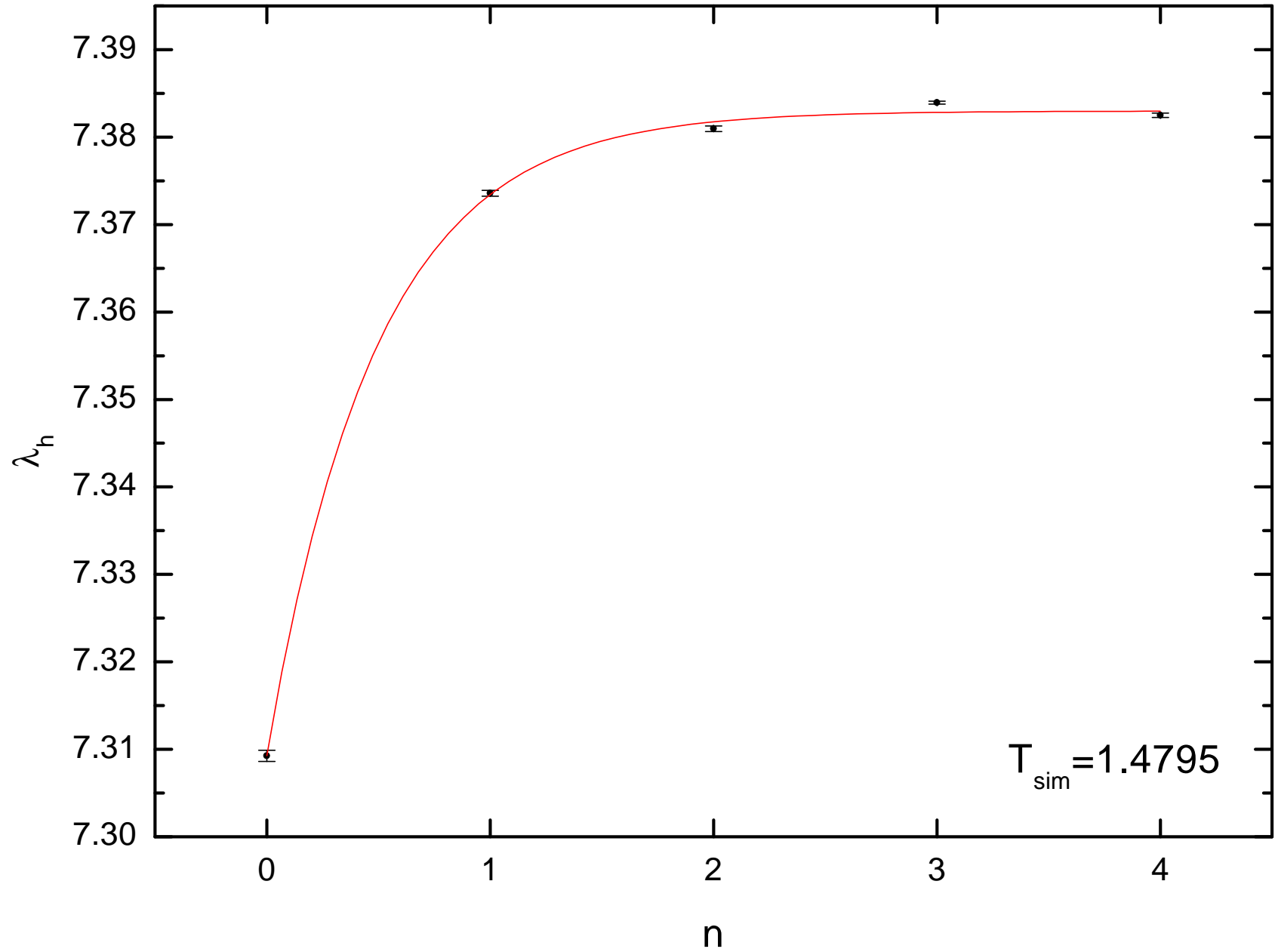
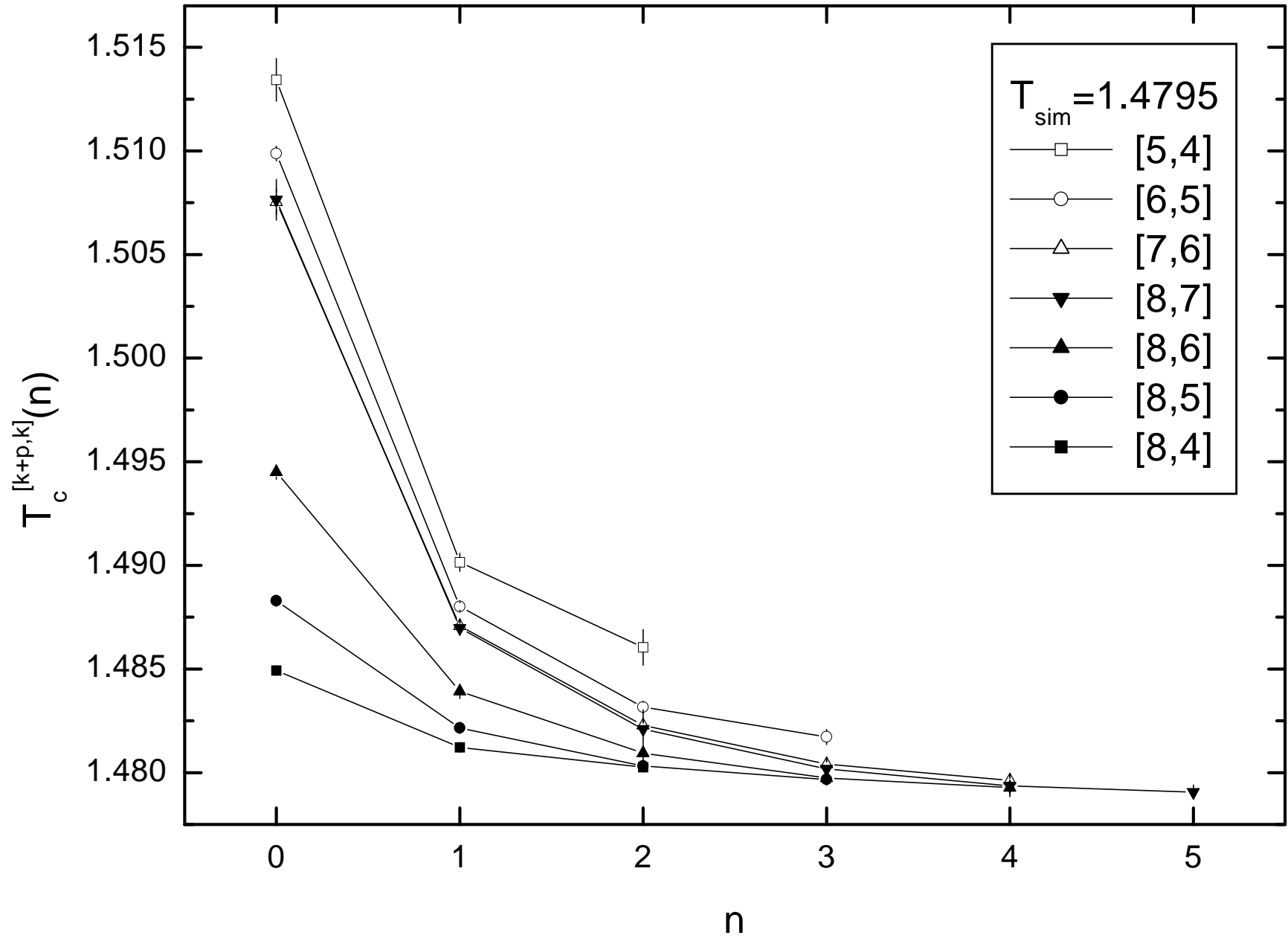
\end{document}